\documentclass[preprint,5p,twocolumn,10pt]{elsarticle}

\usepackage[pdftex,hypertexnames=false,linktocpage=true]{hyperref}
\hypersetup{pdfauthor={Lin Bolang},colorlinks=true,linkcolor=red,anchorcolor=darkgreen,citecolor=green!50!blue,filecolor=darkgreen,urlcolor=green,bookmarksnumbered=true,pdfview=FitB}

\usepackage{lipsum}
\usepackage{mathtools,cuted}

\usepackage{amsmath}

\usepackage{graphicx}
\usepackage{float}
\usepackage{subfigure}

\usepackage{array}
\usepackage{booktabs}
\usepackage{arydshln}
\usepackage{multirow}

\newcommand{\be}{\begin{equation}}  
\newcommand{\ee}{\end{equation}}  

\newcommand{\ba}{\begin{align}}  
\newcommand{\ea}{\end{align}}

\hypersetup{colorlinks=true, citecolor=blue, urlcolor=blue, linkcolor=blue}
\usepackage{braket}

\usepackage{comment}
\begin{document}

\begin{frontmatter}


\title{Skewed generalized parton distributions of proton from basis light-front quantization}

\author[imp,ucas,keylab]{Yiping~Liu}
\ead{liuyiping@impcas.ac.cn}

\author[imp,ucas,keylab]{Siqi~Xu}
\ead{xsq234@impcas.ac.cn}

\author[imp,ucas,keylab]{Chandan~Mondal\corref{cor1}}
\ead{mondal@impcas.ac.cn}

\author[imp,ucas,keylab]{Xingbo~Zhao}
\ead{xbzhao@impcas.ac.cn}

\author[iowa]{James~P.~Vary}
\ead{jvary@iastate.edu}

\author[]{\\\vspace{0.2cm}(BLFQ Collaboration)}

\address[imp]{Institute of Modern Physics, Chinese Academy of Sciences, Lanzhou, Gansu, 730000, China}
\address[ucas]{School of Nuclear Physics, University of Chinese Academy of Sciences, Beijing, 100049, China}
\address[keylab]{CAS Key Laboratory of High Precision Nuclear Spectroscopy, Institute of Modern Physics, Chinese Academy of Sciences, Lanzhou 730000, China}
\address[iowa]{Department of Physics and Astronomy, Iowa State University, Ames, IA 50011, USA}
 \cortext[cor1]{Corresponding author}

\begin{abstract}
We obtain all the leading-twist quark generalized parton distributions (GPDs) inside the proton at nonzero skewness within the basis light-front quantization framework. We employ the light-front wave functions of the proton from a light-front quantized Hamiltonian in the valence Fock sector consisting of a three-dimensional confinement potential and a one-gluon exchange interaction with fixed coupling. We find that the qualitative behaviors of our GPDs are similar to those of other theoretical
calculations. We further examine the GPDs within the boost-invariant longitudinal coordinate, $\sigma=\frac{1}{2} b^- P^+$, which is  identified as the Fourier conjugate of the skewness.
The GPDs in the $\sigma$-space show diffraction patterns, which are akin to the diffractive scattering of a wave in optics.
\end{abstract}
\end{frontmatter}

\section{Introduction}\label{intro}
For more than two decades, extensive experimental and theoretical efforts~(see Ref.~\cite{Diehl:2023nmm} and the references therein) aim to gain insights into the generalized parton distributions (GPDs), which encode the three-dimensional (3D) information of the quark in the proton. In addition to the longitudinal momentum fraction ($x$), the GPDs also depend on the longitudinal momentum transferred ($\xi$),  also  known as the skewness, and the square of the total momentum transferred ($t$). In the forward limit, $t=0$ and $\xi=0$, the GPDs reduce to the ordinary parton distribution functions (PDFs), accessible in the deep inelastic scattering (DIS) process. The Mellin moments of GPDs correspond to the form factors.  The Fourier transform of GPDs at zero skewness, with respect to the transverse momentum transfer yields the impact parameter dependent parton distribution functions~\cite{Burkardt:2000za,Burkardt:2002hr}, which exhibit how a parton with particular longitudinal momentum is distributed in the transverse position space. These distributions adhere to certain positivity conditions and, unlike the GPDs in momentum space, have a probabilistic interpretation~\cite{Ralston:2001xs}. For nonzero skewness, the GPDs can also be represented in the longitudinal position space by performing
Fourier transformation of the GPDs with respect to the skewness variable $\xi$~\cite{Brodsky:2006in,Brodsky:2006ku,Chakrabarti:2008mw,Manohar:2010zm,Kumar:2015fta,Mondal:2015uha,Chakrabarti:2015ama,Mondal:2017wbf}.

Precise knowledge of GPDs is needed for the analysis and interpretation of the exclusive  scattering processes, such as deeply virtual Compton scattering (DVCS)~\cite{Ji:1996nm,Goeke:2001tz},  deeply virtual meson productions (DVMP)~\cite{Goloskokov:2007nt,Collins:1996fb}, timelike Compton scattering (TCS)~\cite{Berger:2001xd}, single diffractive hard exclusive processes (SDHEPs)~\cite{Qiu:2022pla,Grocholski:2022rqj,Duplancic:2022ffo}, wide-angle Compton scattering (WACS)~\cite{Radyushkin:1998rt,Diehl:1998kh}, and also double DVCS~\cite{Deja:2023ahc}. Various experimental collaborations and facilities worldwide, for example, ZEUS~\cite{ZEUS:1998xpo, ZEUS:2003pwh}, H1~\cite{H1:2001nez, H1:2005gdw},  HERMES~\cite{HERMES:2001bob, HERMES:2006pre, HERMES:2008abz} at DESY, Hall A~\cite{JeffersonLabHallA:2006prd, JeffersonLabHallA:2007jdm}, CLAS~\cite{CLAS:2001wjj, CLAS:2006krx, CLAS:2007clm} at Jefferson Lab (JLab), COMPASS~\cite{dHose:2004usi} at CERN, etc., have made considerable efforts to gain insights into GPDs. Recently, GPDs have been obtained by analysing the world electron scattering data~\cite{Hashamipour:2021kes, Hashamipour:2020kip}.
The GPDs give us with vital information about the spatial distributions, spin and orbital motion of quarks inside the proton. Therefore, precise determination of the GPDs of the proton becomes one of the major goals at the upcoming Electron-Ion-Colliders (EICs)~\cite{Accardi:2012qut,AbdulKhalek:2021gbh,Anderle:2021wcy}, the Large
Hadron-Electron Collider (LHeC)~\cite{LHeCStudyGroup:2012zhm,LHeC:2020van}, and the $12$ GeV upgrade program at JLab~\cite{Dudek:2012vr, Burkert:2018nvj}.

Besides experiments, theoretical investigations have also made significant progress. Several theoretical studies for the proton GPDs have been reported that use different QCD inspired models such as the bag model~\cite{Ji:1997gm,Anikin:2001zv}, soliton model \cite{Goeke:2001tz,Petrov:1998kf,Penttinen:1999th}, constituent quark model (CQM) \cite{Scopetta:2003et,Scopetta:2002xq,Boffi:2002yy,Boffi:2003yj,Scopetta:2004wt}, light-front quark-diquark model~\cite{Mondal:2017wbf,Maji:2017ill,Chakrabarti:2015ama,Maji:2015vsa,Mondal:2015uha}, meson cloud model~\cite{Pasquini:2006dv,Pasquini:2006ib}, Anti-de Sitter/QCD~\cite{Vega:2010ns,Chakrabarti:2013gra,Rinaldi:2017roc,Traini:2016jko,deTeramond:2018ecg,Gurjar:2022jkx}, light-front quantization~\cite{Tiburzi:2001ta,Tiburzi:2001je,Mukherjee:2002xi,Lin:2023ezw,Zhang:2023xfe}, etc. Meanwhile, promising theoretical
frameworks for calculating GPDs also include the  Euclidean lattice  QCD~\cite{Ji:2013dva,Ji:2020ect,Lin:2021brq,Lin:2020rxa,Bhattacharya:2022aob, Alexandrou:2021bbo, Alexandrou:2022dtc, Guo:2022upw, Alexandrou:2020zbe, Gockeler:2005cj, QCDSF:2006tkx, Alexandrou:2019ali}.

In this work, we investigate all the leading-twist quark GPDs at nonzero skewness of the proton within the recently developed nonperturbative framework, known as basis light-front quantization (BLFQ)~\cite{Vary:2009gt,Zhao:2014xaa,Nair:2022evk,Lan:2019vui,Mondal:2019jdg,Xu:2021wwj,Lan:2021wok,Xu:2023nqv}. Previously, this approach has been successfully applied to explore the proton GPDs at zero skewness~\cite{Kaur:2023lun}. However, the experiments always probe at $\xi\ne 0$. Therefore, it is important to investigate the GPDs at nonzero skewness from the perspective of theory. Here, we adopt a light-front effective Hamiltonian for the proton in the constituent valence quark Fock space and solve for its mass eigenstates and light-front wave functions (LFWFs). Parameters in this Hamiltonian have been fixed to generate the proton mass and the flavor form factors~\cite{Mondal:2019jdg,Xu:2021wwj}. The resulting LFWFs  have been successfully employed to study various proton properties, e.g., the electromagnetic and axial form factors, radii, PDFs, TMDs, angular momentum distributions, etc.~\cite{Mondal:2019jdg,Xu:2021wwj,Liu:2022fvl,Hu:2022ctr}. Here, we extend those investigations to study the quark GPDs at nonzero skewness.

\section{Proton LFWFs from a light-front effective Hamiltonian in the BLFQ framework \label{Sec2}}
The LFWFs of the proton are obtained from the light-front Hamiltonian by solving the eigenvalue equation: 
$
H_{\rm LF}\vert \Psi\rangle=M^2\vert \Psi\rangle,
$
where $H_{\rm LF}$ represents the  Hamiltonian of the proton with its mass squared ($M^2$) eigenvalue. With quarks being the only explicit degree of freedom, the effective light-front Hamiltonian we employ for the proton is given by~\cite{Mondal:2019jdg}
\begin{align}\label{hami}
&H_{\rm eff}=\sum_a \frac{{\vec k}_{\perp a}^2+m_{a}^2}{x_a}\nonumber\\
&+\frac{1}{2}\sum_{a\ne b}\kappa^4 \Big[x_ax_b({ \vec r}_{\perp a}-{ \vec r}_{\perp b})^2-\frac{\partial_{x_a}(x_a x_b\partial_{x_b})}{(m_{a}+m_{b})^2}\Big]
\\&+\frac{1}{2}\sum_{a\ne b} \frac{C_F 4\pi \alpha_s}{Q^2_{ab}} \bar{u}(k'_a,s'_a)\gamma^\mu{u}(k_a,s_a)\bar{u}(k'_b,s'_b)\gamma^\nu{u}(k_b,s_b)g_{\mu\nu}\,,\nonumber
\end{align}
where $x_a$ and ${\vec k}_{\perp a}$ correspond to the longitudinal momentum fraction and the relative transverse momentum of the quark $a$. $m_{a}$ defines the mass of the quark a, and $\kappa$ is the strength of the 3D-confinement, the second term of the Hamiltonian. The variable $\vec{r}_\perp={ \vec r}_{\perp a}-{ \vec r}_{\perp b}$ represents the transverse separation between two quarks. The last term in the Hamiltonian denotes  the one-gluon exchange (OGE) interaction with $Q^2_{ab}=-q^2=-(1/2)(k'_a-k_a)^2-(1/2)(k'_b-k_b)^2$ being the average momentum transfer squared and  $\alpha_s$ is the coupling constant, $C_F =-2/3$ corresponds to the color
factor, and $g_{\mu\nu}$ defines the metric tensor. ${u}(k_a,s_a)$ denotes the spinor with momentum $k_a$ and spin $s_a$. The explicit expression of the OGE has been derived in Refs~\cite{Wiecki:2014ola,Li:2015zda}.

In the BLFQ framework~\cite{Vary:2009gt}, we adopt the two-dimensional harmonic oscillator (2D-HO) functions to describe transverse degrees of freedom and the discretized plane-wave basis in the longitudinal direction~\cite{Zhao:2014xaa}. Diagonalizing the Hamiltonian, Eq.~(\ref{hami}), in our chosen basis space provides the eigenvalues as mass squares of the hadronic bound states, and the eigenfunctions corresponding to the LFWFs that encode the structure of bound
states. 
The ground-state is naturally identified as the proton state defined as $\ket{P, {\Lambda}}$, with $P$ and $\Lambda$ being the momentum and the helicity of the state. The LFWFs of the proton in terms of the basis functions are expressed as 
\begin{align}
\Psi^{\Lambda}_{\{x_i,\vec{k}_{i\perp},\lambda_i\}}=\sum_{\{n_i,m_i\}} \psi^{\Lambda}_{\{x_{i},n_{i},m_{i},\lambda_i\}} \prod_i \phi_{n_i,m_i}(\vec{k}_{i\perp};b) \,,\label{wavefunctions}
\end{align}
with $\psi^{\Lambda}_{\{x_{i},n_{i},m_{i},\lambda_i\}}=\braket{P, {\Lambda}|\{x_i,n_i,m_i,\lambda_i\}}$ being the amplitudes of the LFWF, expressed in the BLFQ basis, generated by diagnalizing Eq.~(\ref{hami}) numerically. 
$\phi_{n,m}(\vec{k}_{\perp};b)$ is the 2D-HO function with $b$ as its scale parameter~\cite{Zhao:2014xaa}; $n$ and $m$ define the principal and orbital quantum
numbers, respectively, and $\lambda$ denotes the quark helicity.  

The longitudinal momentum fraction $x$, in the discretized plane-wave basis, is defined as
$
x_i=p_i^+/P^+=k_i/K,
$
where $k=\frac{1}{2}, \frac{3}{2}, \frac{5}{2}, ...$, signifying the choice of antiperiodic boundary conditions, and $K=\sum_i k_i$. The transverse basis truncation is specified by $N_{\rm max}$, such that $\sum_i (2n_i+| m_i |+1)\le N_{\rm max}$. The basis cutoff $N_{\rm max}$ acts implicitly as a regulator for the LFWFs in the transverse direction. The corresponding ultraviolet (UV) and infrared (IR)  cutoffs are given by $\Lambda_{\rm UV}\approx b \sqrt{N_{\rm max}}$ and  $\Lambda_{\rm IR}\approx b /\sqrt{N_{\rm max}}$, respectively~\cite{Zhao:2013cma}. The longitudinal basis cutoff $K$  represents the resolution in the
longitudinal direction. 
%
In this paper, all the calculations are performed with $ N_{\rm max}=10, K=16.5 $ and the parameters in the effective Hamiltonian are determined to generate the proton mass and to fit the flavor form factors~\cite{Xu:2021wwj}. 
The LFWFs in this model, suitable for low-resolution probe $\mu_0^2\sim 0.195$ GeV$^2$~\cite{Mondal:2019jdg}, have been successfully applied to compute a wide class of different and related proton observables, e.g., the electromagnetic and axial form factors, radii, PDFs, helicity asymmetries, TMDs, etc., with remarkable overall success~\cite{Mondal:2019jdg,Xu:2021wwj,Liu:2022fvl,Kaur:2023lun,Zhang:2023xfe,Hu:2022ctr}.

\section{Generalized parton distributions \label{Sec3}}
At the leading-twist, there are eight quark GPDs for the proton. Four of them ($H$, $E$, $\tilde{H}$, 
 and $\tilde{E} $) are chirally even  and other four GPDs ($H_T$, $E_T$ $\tilde{H}_T$, and  $\tilde{E}_T$) are chirally odd. They are defined through the off-forward matrix elements of the bilocal operators between the proton states: 
 \begin{align}
F^{[\Gamma]}_{\Lambda\Lambda^\prime}=\int\frac{dy^-}{8\pi}e^{ixP^+y^-/2}\langle P^\prime,\Lambda^\prime|\bar{\psi}(0)\Gamma\psi(y)|P,\Lambda\rangle|_{y^+=y_\perp=0},
\end{align}
with $\Gamma$ being the leading-twist Dirac $\gamma$-matrices, i.e., $\Gamma=\{\gamma^+,\, \gamma^+\gamma^5,\, i\sigma^{j+} \gamma^5\}$ corresponding to unpolarized, longitudinally polarized and transversely polarized quarks, respectively. In the symmetric frame, the average momentum of proton $\bar{P}= \frac{1}{2} (P^{\prime}+P)$, while momentum transfer $\Delta=(P^{\prime}-P)$. The initial and final four momenta of the proton are then given by
\begin{align}
P &\equiv \left((1+\xi)\bar{P}^+,\frac{M^2+\Delta_\perp^2/4}{(1+\xi)P^+},-\vec{\Delta}_\perp/2\right),\label{Pp}\\
P^{\prime} &\equiv \left((1-\xi)\bar{P}^+,\frac{M^2+\Delta_\perp^2/4}{(1-\xi)P^+},\vec{\Delta}_\perp/2\right). \label{Ppp}
\end{align}
The GPDs are then parameterized as~\cite{Diehl:2003ny}
\begin{align}
F^{[\gamma^+]}_{\Lambda\Lambda^\prime}=&\frac{1}{2\bar{P}^+}\bar{u}(P^\prime,\Lambda^\prime)\Big[\gamma^+H+\frac{i\sigma^{+i}\Delta_i}{2M}E\Big]u(P,\Lambda),\nonumber\\
F^{[\gamma^+\gamma_5]}_{\Lambda\Lambda^\prime}=&\frac{1}{2\bar{P}^+}\bar{u}(P^\prime,\Lambda^\prime)\Big[\gamma^+\gamma_5\tilde{H}+\frac{\Delta^+\gamma_5}{2M}\tilde{E}\Big]u(P,\Lambda),\nonumber\\
F^{[i\sigma^{j+}\gamma_5]}=&-\frac{i\epsilon^{ij}_T}{2\bar{P}^+}\bar{u}(P^\prime,\Lambda^\prime)\Big[i\sigma^{+i}H_T\nonumber\\
&+\frac{\gamma^+\Delta_i-\Delta^+\gamma_i}{2M}E_T+\frac{\bar{P}^+\Delta_i-\Delta^+\bar{P}_i}{M^2}\tilde{H}_T\nonumber\\
&+\frac{\gamma^+\bar{P}_i-\bar{P}^+\gamma_i}{M}\tilde{E}_T\Big]u(P,\Lambda),
\end{align}
where $\epsilon^{ij}_T=\epsilon^{+-ij}$ with the antisymmetric Levi-Civita tensor $\epsilon^{+-12}=1$. Here, $i$ and $j$ can only be transverse indices $1,\,2$. Note that all the GPDs are functions of $x$, $\xi=- \Delta^+/2\bar{P}^+$, and $t= \Delta^2$. One can derive the following relation explicitly from $\Delta^-$
\begin{align}
- t= \frac{4 \xi^2 M^2 + \Delta_\perp^2}{(1-\xi^2)}\,. \label{mt_def}
\end{align}

Using the reference frame where the momenta $\vec{P}_\perp^{\prime}$ and $\vec{P}_\perp$ lie in the $x-z$ plane, we explicitly derive the following
relations for the chiral-even GPDs
\begin{equation}
\begin{aligned}
&H(x,\xi,t)=\frac{1}{\sqrt{1-\xi^2}}F^{[\gamma^+]}_{++}+\frac{2M\xi^2}{\sqrt{1-\xi^2}\Delta_{\perp 1}}F^{[\gamma^+]}_{-+},\\&E(x,\xi,t)=\frac{2M\sqrt{1-\xi^2}}{\Delta_{\perp 1}}F^{[\gamma^+]}_{-+},\\&\tilde{H}(x,\xi,t)=\frac{1}{\sqrt{1-\xi^2}}F^{[\gamma^+\gamma_5]}_{++}+\frac{2M\xi}{\sqrt{1-\xi^2}\Delta_{\perp 1}}F^{[\gamma^+\gamma_5]}_{-+},\\&\tilde{E}(x,\xi,t)=-\frac{2M\sqrt{1-\xi^2}}{\xi\Delta_{\perp 1}}F^{[\gamma^+\gamma_5]}_{-+},
\end{aligned}
\end{equation}
where the proton helicity  is designated by $\Lambda=+(-)$, corresponding to $+1(-1)$, respectively.
For the chiral-odd GPDs, we transform the matrix elements from the helicity basis to the transversity basis~\cite{Pasquini:2005dk,Chakrabarti:2015ama} and decompose $i\sigma^{1+}\gamma_5$ into $\gamma^+\gamma^1\gamma_5$ and $\frac{i}{2}\sigma^{+1}$. In the transversity basis, we obtain
\begin{equation}
\begin{aligned}
H_T(x,\xi,t)=&\frac{1}{\sqrt{1-\xi^2}}F^{[\gamma^+\gamma^1\gamma_5]}_{\uparrow\uparrow}\\
&+\frac{2M\xi}{\sqrt{1-\xi^2}\Delta_{\perp 1}}F^{[\gamma^+\gamma^1\gamma_5]}_{\downarrow\uparrow},\\E_T(x,\xi,t)=&-\frac{2M}{\sqrt{1-\xi^2}\Delta_{\perp 1}}\Big(\xi F^{[\gamma^+\gamma^1\gamma_5]}_{\downarrow\uparrow}+F^{[\frac{i}{2}\sigma^{+1}]}_{\uparrow\uparrow}\Big)\\&-\frac{4M^2}{\sqrt{1-\xi^2}\Delta_{\perp 1}^2}\Big(F^{[\frac{i}{2}\sigma^{+1}]}_{\uparrow\downarrow}-F^{[\gamma^+\gamma^1\gamma_5]}_{\uparrow\uparrow}\Big),\\
\tilde{H}_T(x,\xi,t)=&\frac{2M^2\sqrt{1-\xi^2}}{\Delta_{\perp 1}^2}\Big(F^{[\frac{i}{2}\sigma^{+1}]}_{\uparrow\downarrow}-F^{[\gamma^+\gamma^1\gamma_5]}_{\uparrow\uparrow}\Big),\\
\tilde{E}_T(x,\xi,t)=&-\frac{2M}{\sqrt{1-\xi^2}\Delta_{\perp 1}}\Big(F^{[\gamma^+\gamma^1\gamma_5]}_{\downarrow\uparrow}+\xi F^{[\frac{i}{2}\sigma^{+1}]}_{\uparrow\uparrow}\Big)\\&-\frac{4M^2\xi}{\sqrt{1-\xi^2}\Delta_{\perp 1}}\Big(F^{[\frac{i}{2}\sigma^{+1}]}_{\uparrow\downarrow}-F^{[\gamma^+\gamma^1\gamma_5]}_{\uparrow\uparrow}\Big),
\end{aligned}
\end{equation}
where $\Lambda_T=\uparrow (\downarrow)$ labels the transverse polarization of the
proton polarized along the +ve ($\uparrow$) or
-ve ($\downarrow$) direction.
In overlap representation, $F^{[\Gamma]}_{\Lambda\Lambda^\prime}$ and $F^{[\Gamma]}_{\Lambda_T\Lambda^\prime_T}$  are expressed in terms of the LFWFs as 
\begin{figure*}[htp]
\includegraphics[scale=.35]{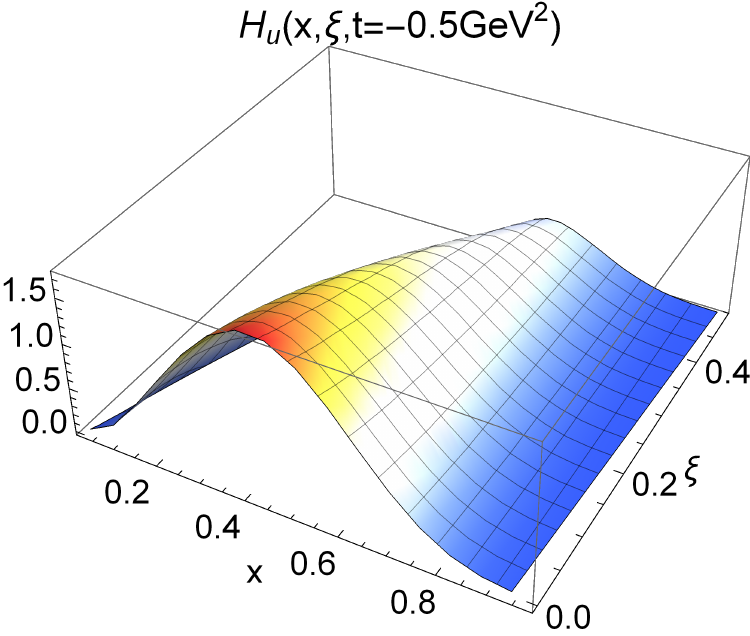}
\includegraphics[scale=.35]{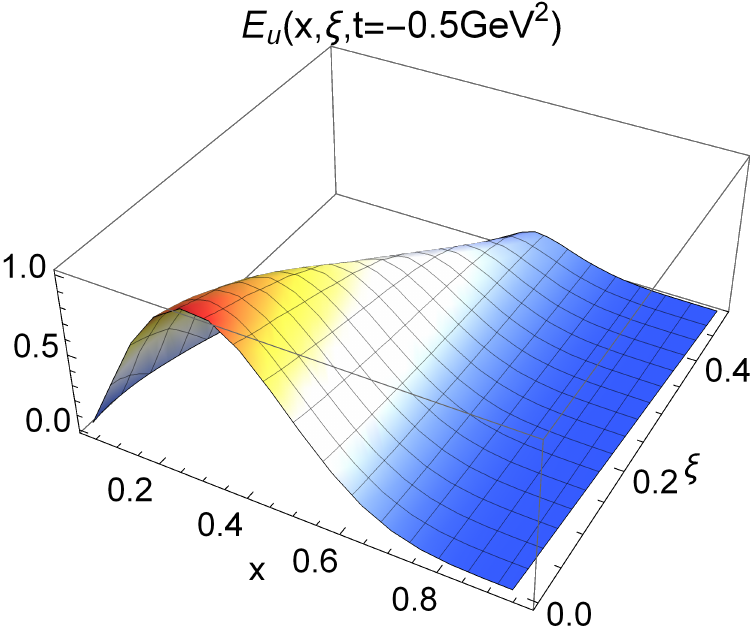} 
\includegraphics[scale=.35]{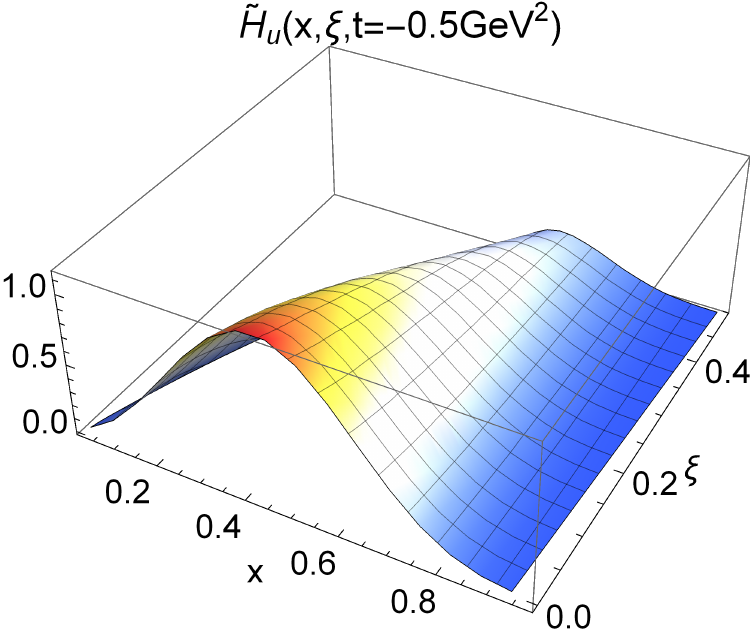} 
\includegraphics[scale=.35]{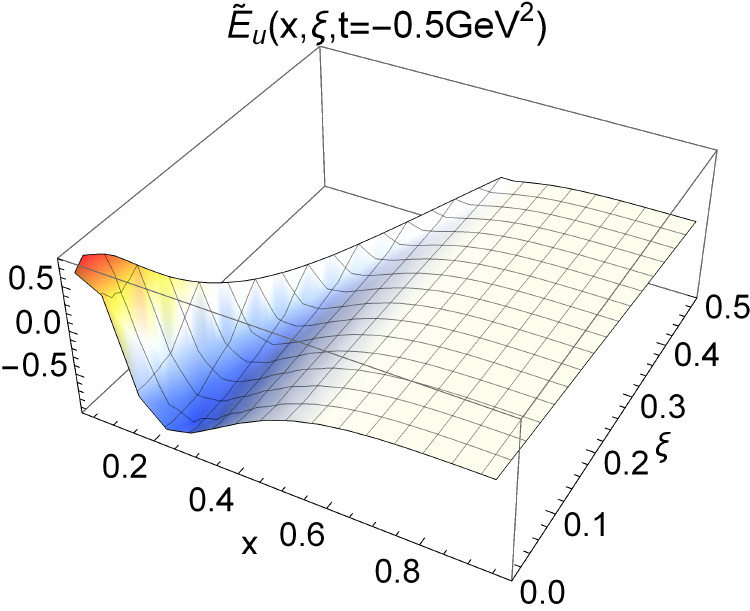} \\
\includegraphics[scale=.35]{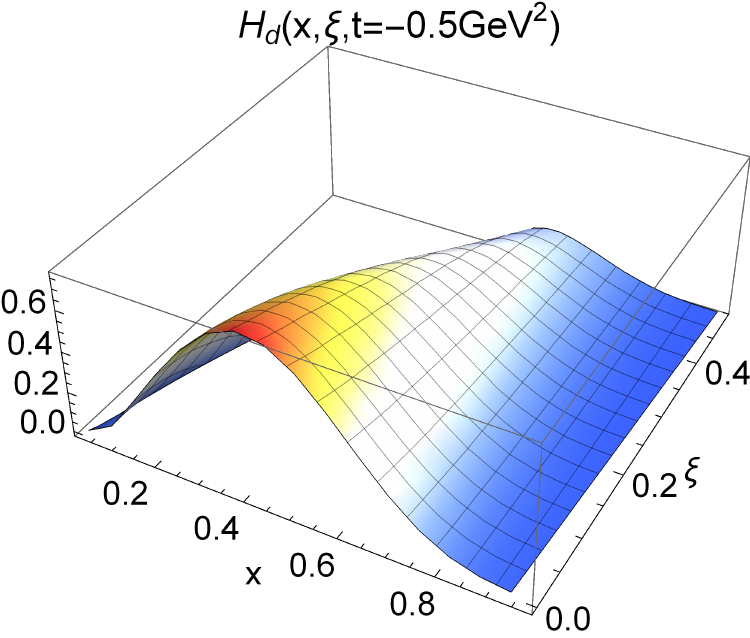}
\includegraphics[scale=.35]{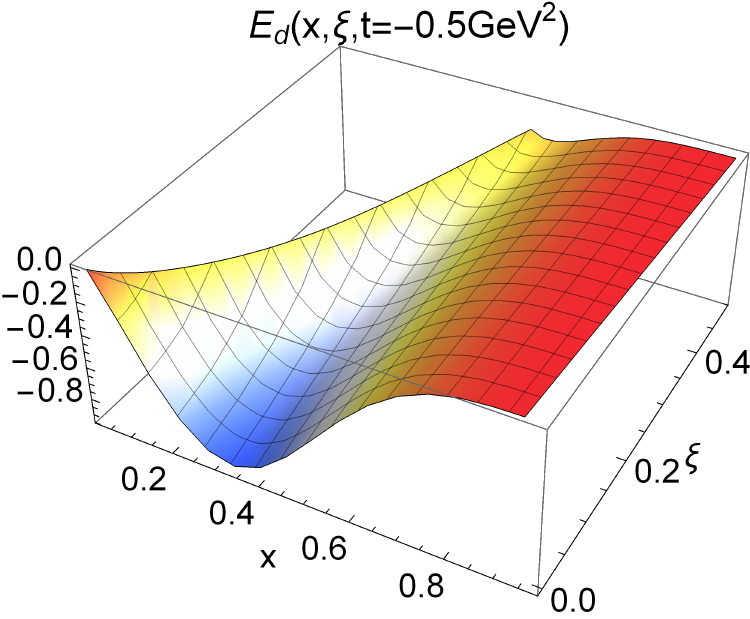} 
\includegraphics[scale=.35]{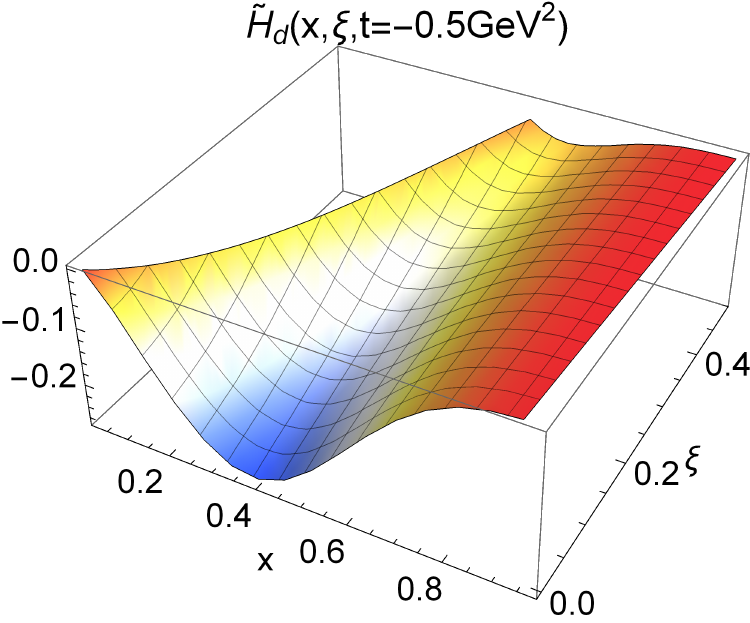} 
\includegraphics[scale=.35]{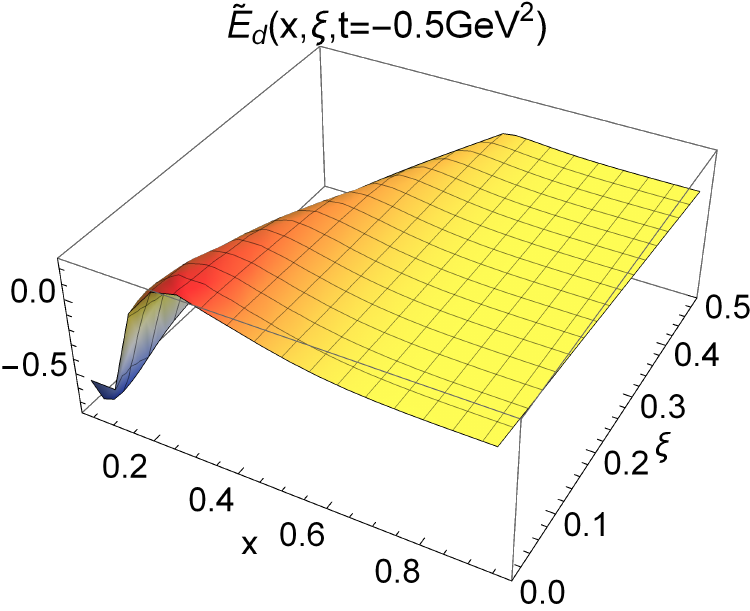}
\caption{\label{chiral_even_xz} The chiral-even GPDs as functions of $x$ and  $\xi$ for fixed $t=-0.5$ GeV$^2$. The upper (lower) panel is for the $u \,(d)$  quark.  Left to right panels represent the GPDs $H,\, E,\, \tilde{H}$, and  $\tilde{E}$, respectively.}
\end{figure*}
\begin{equation}
\begin{aligned}
F^{[\gamma^+]}_{\Lambda\Lambda^\prime}=&\sum_{\{\lambda_i\}} \int \left[{\rm d}\mathcal{X} \,{\rm d}\mathcal{P}_\perp\right]\, \Psi^{\Lambda^\prime *}_{\{x^{\prime\prime}_i,\vec{k}^{\prime\prime}_{i\perp},\lambda_i\}}\Psi^{\Lambda}_{\{x_i^{\prime},\vec{k}_{i\perp}^{\prime},\lambda_i\}},
\\
F^{[\gamma^+\gamma_5]}_{\Lambda\Lambda^\prime}=& \sum_{\{\lambda_i\}} \int \left[{\rm d}\mathcal{X} \,{\rm d}\mathcal{P}_\perp\right]\, \lambda_1\,\Psi^{\Lambda^\prime *}_{\{x^{\prime\prime}_i,\vec{k}^{\prime\prime}_{i\perp},\lambda_i\}}\Psi^{\Lambda}_{\{x_i^{\prime},\vec{k}_{i\perp}^{\prime},\lambda_i\}},
\\
F^{[\gamma^+\gamma_1\gamma_5]}_{\uparrow\uparrow}=& \sum_{\{\lambda_i,\,\lambda_i^\prime\}} \int \left[{\rm d}\mathcal{X} \,{\rm d}\mathcal{P}_\perp\right]\, \Psi^{- *}_{\{x^{\prime\prime}_i,\vec{k}^{\prime\prime}_{i\perp},\lambda_i^\prime\}}\Psi^{+}_{\{x_i^{\prime},\vec{k}_{i\perp}^{\prime},\lambda_i\}} \\&\times\delta_{\lambda_1^\prime,-\lambda_1}\delta_{\lambda_{2,3}^\prime,\lambda_{2,3}},
\\
F^{[\gamma^+\gamma_1\gamma_5]}_{\downarrow\uparrow}=& \sum_{\{\lambda_i,\,\lambda_i^\prime\}} \int \left[{\rm d}\mathcal{X} \,{\rm d}\mathcal{P}_\perp\right]\, \Psi^{+ *}_{\{x^{\prime\prime}_i,\vec{k}^{\prime\prime}_{i\perp},\lambda_i^\prime\}}\Psi^{+}_{\{x_i^{\prime},\vec{k}_{i\perp}^{\prime},\lambda_i\}} \\&\times \lambda_1^\prime \,\delta_{\lambda_1^\prime,-\lambda_1}\delta_{\lambda_{2,3}^\prime,\lambda_{2,3}},
\\
F^{[\frac{i}{2}\sigma^{+1}]}_{\uparrow\uparrow}=& \sum_{\{\lambda_i,\,\lambda_i^\prime\}} \int \left[{\rm d}\mathcal{X} \,{\rm d}\mathcal{P}_\perp\right]\, \Psi^{+ *}_{\{x^{\prime\prime}_i,\vec{k}^{\prime\prime}_{i\perp},\lambda_i^\prime\}}\Psi^{+}_{\{x_i^{\prime},\vec{k}_{i\perp}^{\prime},\lambda_i\}} \\&\times \delta_{\lambda_1^\prime,-\lambda_1}\delta_{\lambda_{2,3}^\prime,\lambda_{2,3}},
\\
F^{[\frac{i}{2}\sigma^{+1}]}_{\downarrow\uparrow}=& \sum_{\{\lambda_i,\,\lambda_i^\prime\}} \int \left[{\rm d}\mathcal{X} \,{\rm d}\mathcal{P}_\perp\right]\, \Psi^{- *}_{\{x^{\prime\prime}_i,\vec{k}^{\prime\prime}_{i\perp},\lambda_i^\prime\}}\Psi^{+}_{\{x_i^{\prime},\vec{k}_{i\perp}^{\prime},\lambda_i\}} \\&\times (-\lambda_1^\prime) \,\delta_{\lambda_1^\prime,-\lambda_1}\delta_{\lambda_{2,3}^\prime,\lambda_{2,3}},
\end{aligned}
\end{equation}
where 
\begin{align}
\left[{\rm d}\mathcal{X} \,{\rm d}\mathcal{P}_\perp\right]=&\prod_{i=1}^3 \left[\frac{{\rm d}x_i{\rm d}^2 \vec{k}_{i\perp}} {16\pi^3}\right]\delta(x-x_1)\nonumber\\
&\times16 \pi^3 \delta \left(1-\sum_{i=1}^{3} x_i\right) \delta^2 \left(\sum_{i=1}^{3}\vec{k}_{i\perp}\right),  
\end{align}
and the light-front momenta are $x^{\prime}_1=\frac{x_1+\xi}{1+\xi}$; $\vec{k}^{\prime}_{1\perp}=\vec{k}_{1\perp}+(1-x^{\prime})\frac{\vec{\Delta}_{\perp}}{2}$ for the initial struck quark ($i=1$) and $x^{\prime}_i=\frac{x_i}{1+\xi}; ~\vec{k}^{\prime}_{i\perp}=\vec{k}_{i\perp}-{x_i^{\prime}} \frac{\vec{\Delta}_{\perp}}{2}$ for the initial spectators ($i\ne1$) and
$x^{\prime\prime}_1=\frac{x_1-\xi}{1-\xi}$; $\vec{k}^{\prime}_{1\perp}=\vec{k}_{1\perp}-(1-x^{\prime\prime})\frac{\vec{\Delta}_{\perp}}{2}$ for the final struck quark and $x^{\prime\prime}_i=\frac{x_i}{1-\xi}; ~\vec{k}^{\prime}_{i\perp}=\vec{k}_{i\perp}+{x_i^{\prime\prime}} \frac{\vec{\Delta}_{\perp}}{2}$ for the final spectators and
$\lambda_1 (\lambda_1^\prime)$ designates the struck quark helicity.
Note that, in this work, we consider only the valence Fock component, and therefore, limit ourselves to the Dokshitzer-Gribov-Lipatov-Altarelli-Parisi (DGLAP) domain, $\xi<x<1$, where the number of quarks in the initial and the final states remains conserved.
\begin{figure*}[htp]
\includegraphics[scale=.35]{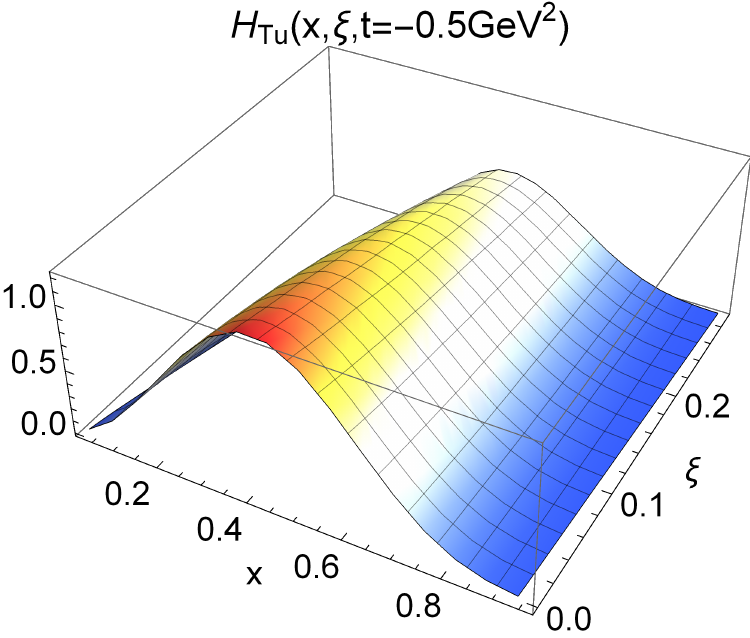}
\includegraphics[scale=.35]{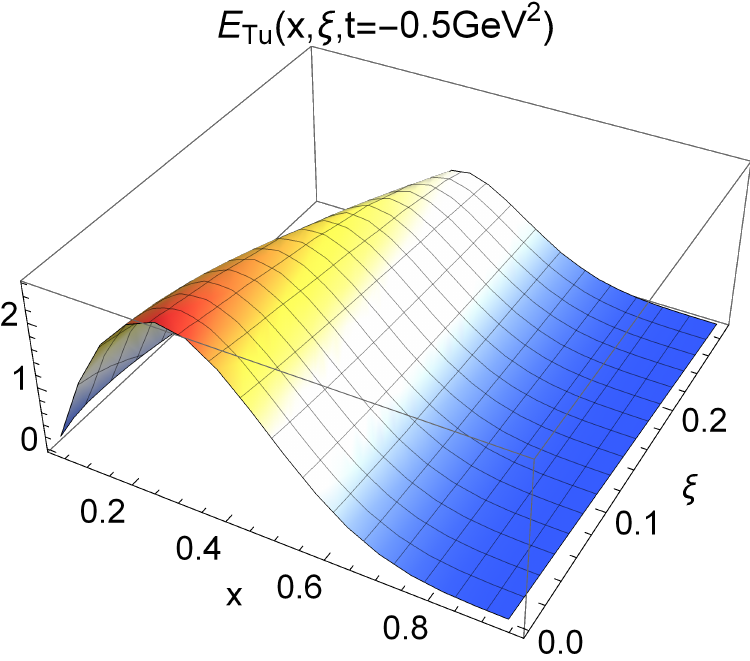} 
\includegraphics[scale=.35]{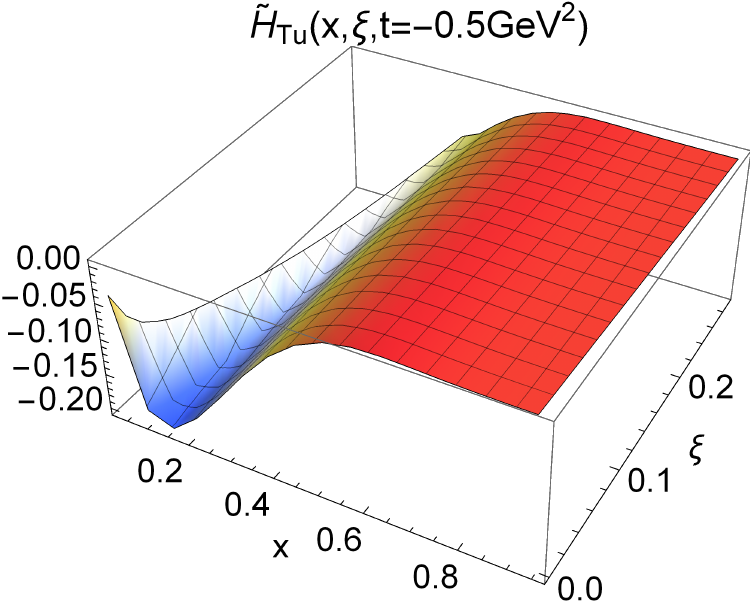} 
\includegraphics[scale=.35]{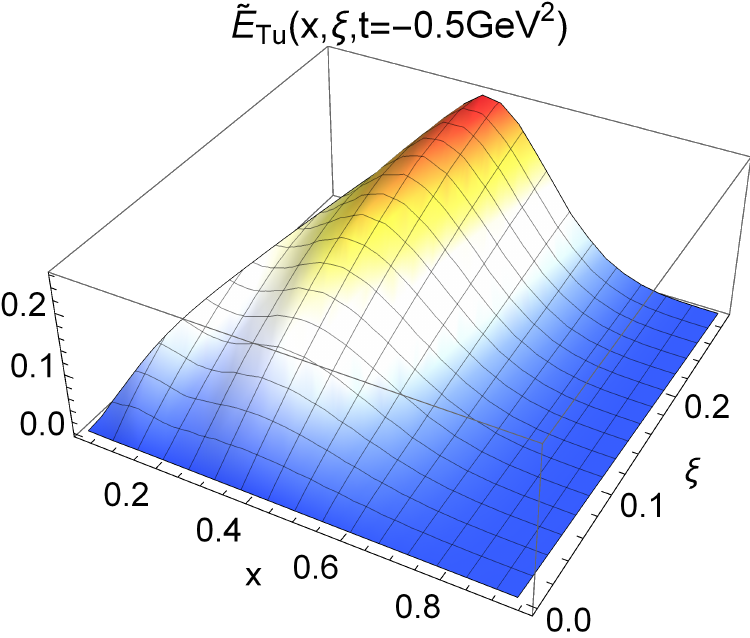} \\
\includegraphics[scale=.35]{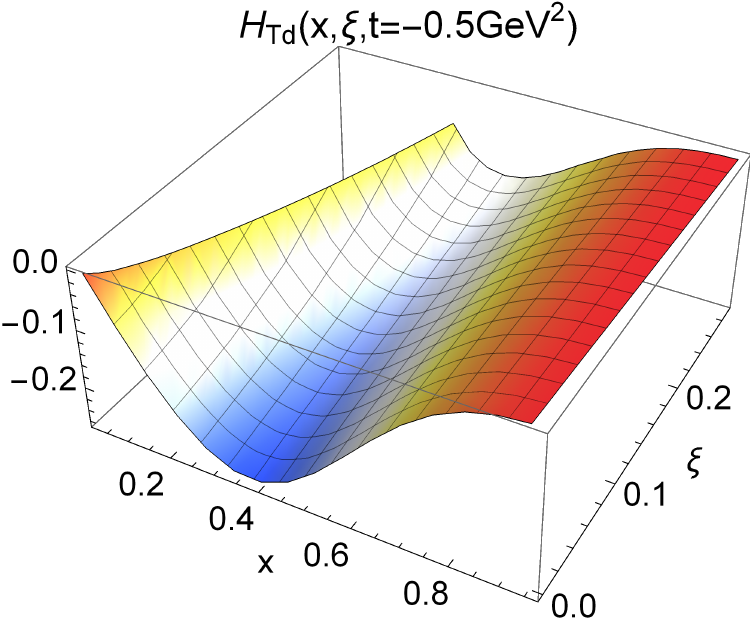}
\includegraphics[scale=.35]{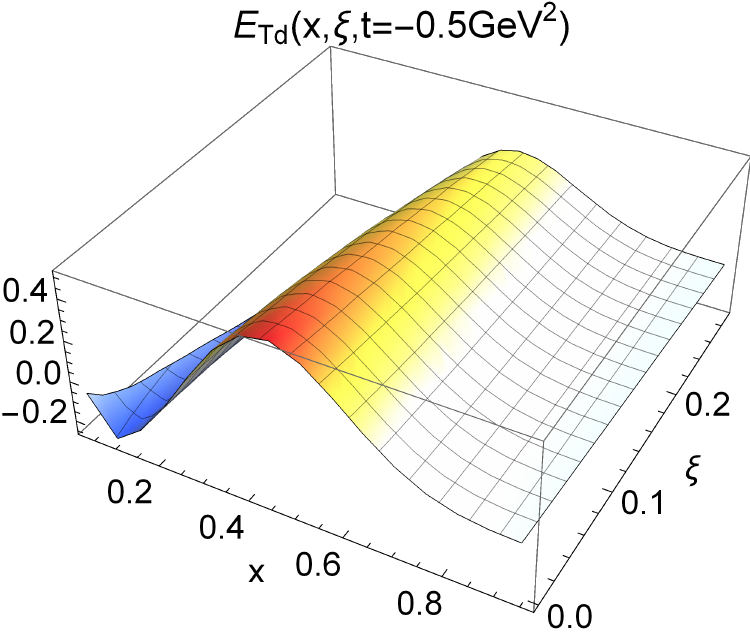} 
\includegraphics[scale=.35]{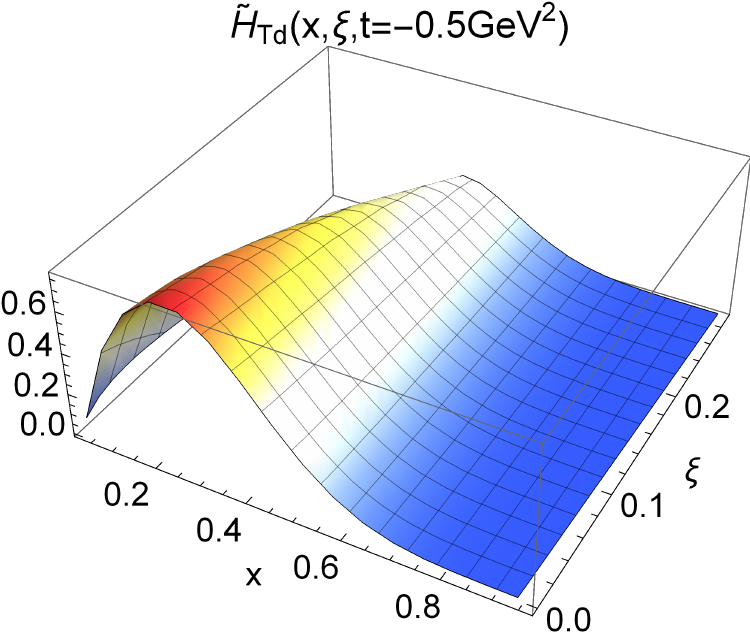} 
\includegraphics[scale=.35]{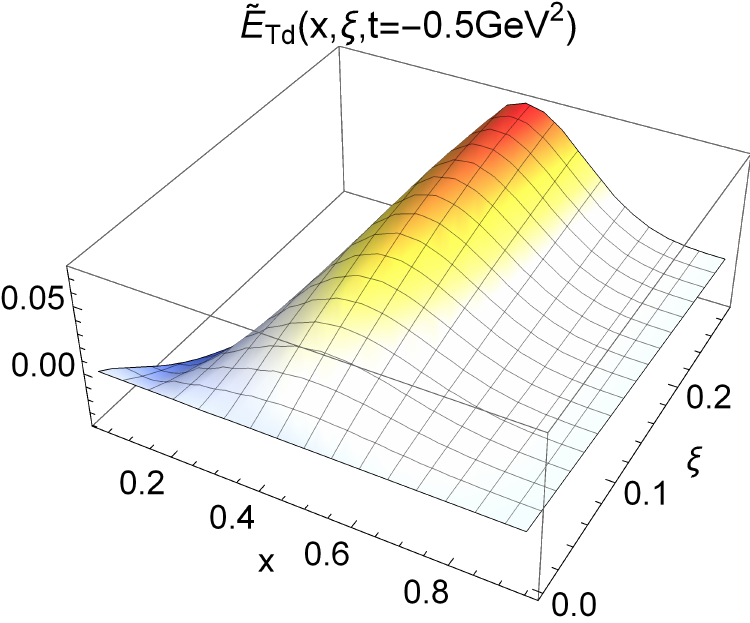}
\caption{\label{chiral_odd_xz} The chiral-odd GPDs as functions of $x$ and $\xi$ for fixed $t=-0.5$ GeV$^2$. The upper (lower) panel is for the $u\, (d)$  quark.   Left to right panels represent the GPDs $H_T,\, E_T,\, \tilde{H}_T$, and  $\tilde{E}_T$, respectively.}
\end{figure*}
\section{Numerical results and discussions}
\label{rsult}
In our BLFQ basis approach, the choice of the discretized plane-wave basis in the longitudinal direction leads to discretized momentum fraction carried by the the quarks, i.e., $x_i=k_i/K$, with $k_i$ being the half-integers. Note that the computation of the  GPDs at nonzero skewness requires the overlap of LFWFs at different longitudinal momentum. We interpolate the longitudinal degree of freedom of our LFWFs and employ them to obtain all the leading-twist GPDs. These GPDs are related to several physical quantities like OAM, axial and tensor charges,  etc., and also linked to the
form factors and the PDFs in certain kinematical limits~\cite{Diehl:2003ny}.
To illustrate the numerical results of the GPDs, we emphasize the $\xi$-dependence of the GPDs since the other dependencies of the GPDs with vanishing skewness have been investigated in several studied~\cite{Liu:2022fvl,Kaur:2023lun}. 

In Fig.~\ref{chiral_even_xz}, we show our results for the chiral-even  GPDs as functions of $x$ and $\xi$ for fixed value of $t=-0.5$ GeV$^2$.  We observe that the general features of many of the GPDs are similar to within an overall sign and scale factor while the $\tilde{E}$ merits discussion (see below). All of these GPDs have their maxima at lower-$x \,(<0.5)$ and the peaks shift towards larger values of $x$ with gradually  decreasing magnitudes as the momentum transfer increases in the longitudinal direction. All the GPDs in the large $x$-region eventually decay and become independent of $\xi$. The chiral-even GPDs: $H,\, E,$ and $\tilde{H}$ for the $u$-quark are positive. Since the anomalous magnetic moment and the axial charge are  negative for the $d$-quark, the related GPDs, $E$ and $\tilde{H}$, are likewise negative. The GPD $\tilde{E}$ exhibits distinctly different behaviour compared to other chiral-even GPDs. We observe $\tilde{E}$ points crossing zero along $x$ and the $u$-quark distribution is opposite to that of the d-quark distribution.
%
We note that the qualitative behaviors of our GPDs are similar to those of other theoretical calculations in Refs.~\cite{Ji:1997gm, Boffi:2002yy, Boffi:2003yj, Mondal:2015uha, Mondal:2017wbf, Freese:2020mcx}.

All the chiral-odd GPDs, presented in Fig.~\ref{chiral_odd_xz}, show similar qualitative behavior as for the chiral-even GPDs, except the behavior of $\tilde{E}_T$. The GPD $\tilde{E}_T$ vanishes for $\xi=0$, since it appears to be an odd function under the transformation $\xi \rightarrow -\xi$. The peak of this distribution shifts towards larger values of $x$ with increasing magnitude as $\xi$ increases. The GPDs $H_T^u$ and $\tilde{H}_T^u$, show opposite behavior to that of the $d$-quark distributions. An interesting distribution arises from the combination of two chiral-odd GPDs, $2\tilde{H}^q_T+E^q_T$, which offers insights into the angular momentum contribution under specific limits and can be reduced to the tensor form factor. Note that all the distributions show the accessibility of the DGLAP region $x>\xi$ and the width of all the distributions in $x$ decrease as $\xi$ increases.  The qualitative behaviors of our chiral-odd GPDs are consistent with other theoretical approaches in Refs.~\cite{Pasquini:2005dk, Chakrabarti:2015ama}.

In the forward limit, the GPDs are reducible to the PDFs, particularly, the unpolarized, helicity-dependent, and transversity, i.e., $H(x,0,0)=f_1(x)$, $\tilde{H}(x,0,0)=g_1(x)$, and $H_T(x,0,0)=h_1(x)$, respectively. Investigations  of these PDFs for valence quarks as well as the electromagnetic, axial, and tensor form factors corresponding to the GPDs $H$ and $E$, $\tilde{H}$, and $2\tilde{H}^q_T+E^q_T$ in our BLFQ approach have been reported in Refs.~\cite{Xu:2021wwj,Kaur:2023lun}.
\begin{figure*}[htp]
\includegraphics[scale=.35]{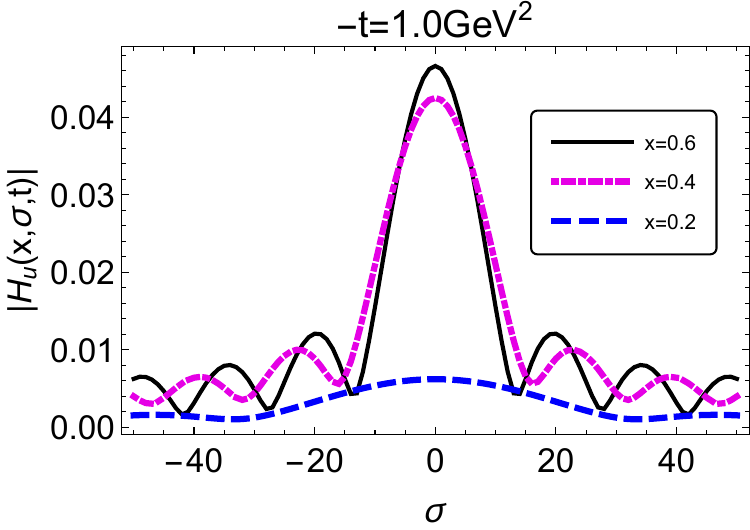}
\includegraphics[scale=.35]{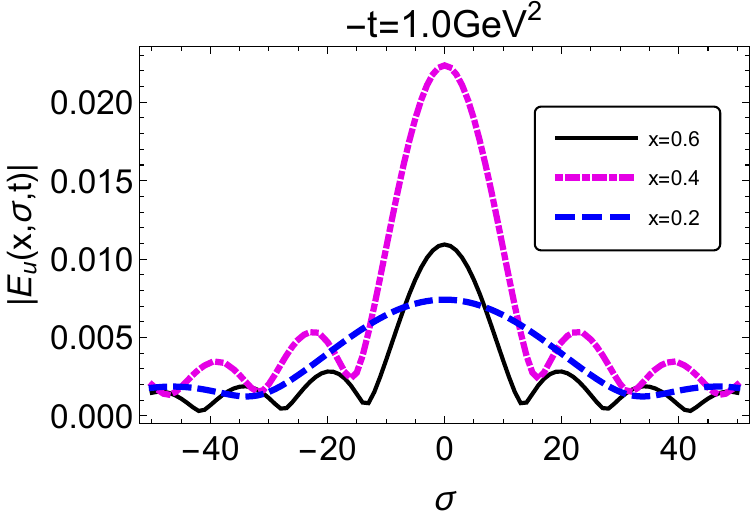} 
\includegraphics[scale=.35]{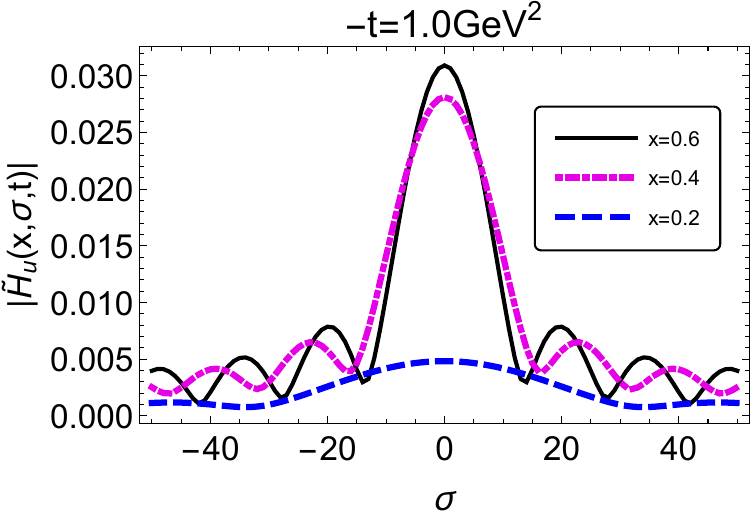} 
\includegraphics[scale=.35]{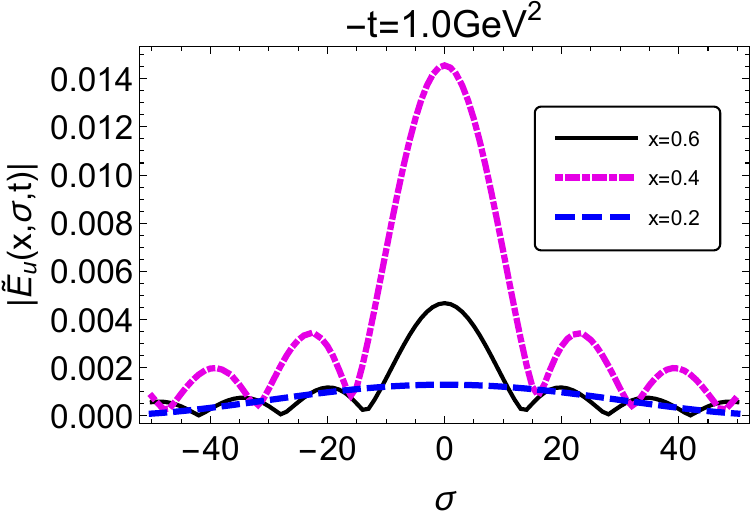} \\
\includegraphics[scale=.35]{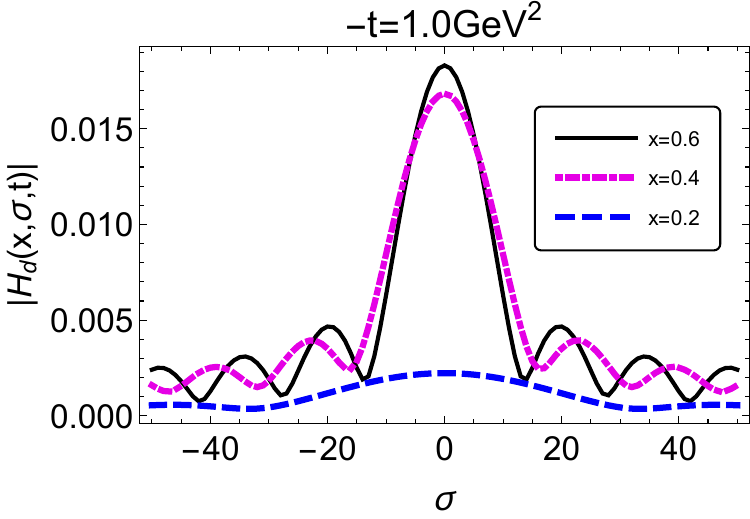}
\includegraphics[scale=.35]{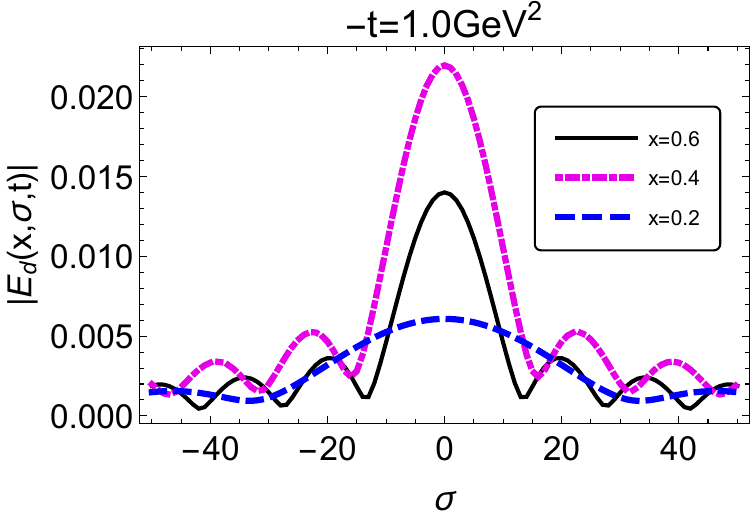} 
\includegraphics[scale=.35]{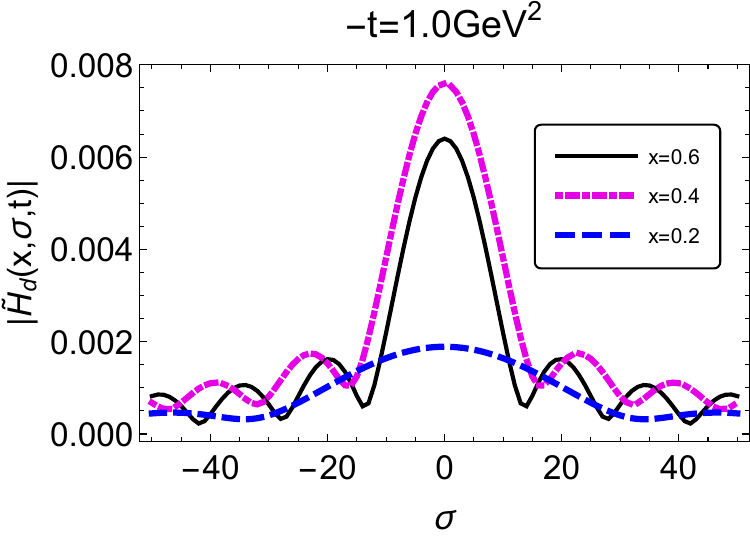} 
\includegraphics[scale=.35]{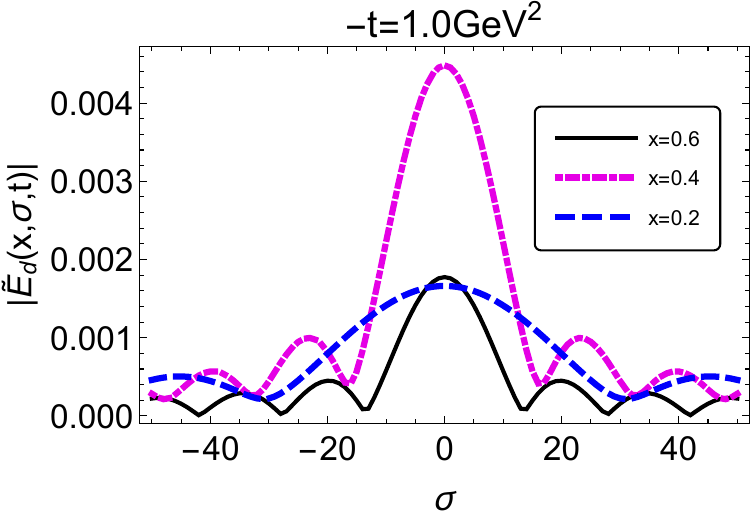}
\caption{\label{chiral_even_xs} The chiral-even GPDs in boost-invariant longitudinal position space as functions of $\sigma$ for  $x=\{0.2,\,0.4,\, 0.6\}$ and fixed $t=-1.0$ GeV$^2$. The upper (lower) panel is for the $u \,(d)$  quark.  Left to right panels represent the GPDs $H,\, E,\, \tilde{H}$, and  $\tilde{E}$, respectively.}
\end{figure*}
\subsection{GPDs in boost-invariant longitudinal space} \label{sec_GPDs_sigma}
The GPDs in the transverse position space have been studied extensively using several theoretical approaches including the BLFQ framework for zero skewness. The transverse impact-parameter $\vec{b}_\perp$ is the Fourier conjugate to the variable $\vec{D}_\perp=\vec{\Delta}_\perp/(1-\xi^2)$~\cite{Diehl:2002he,Burkardt:2002hr,Ralston:2001xs,Kaur:2018ewq}, which simply reduces to $\vec{\Delta}_\perp$ for zero skewness.
Similarly, the longitudinal momentum transfer, $\xi P^+$, can be identified as the Fourier conjugate to the longitudinal distance $\frac{1}{2}b^-$. In other words, $\xi$ is conjugate to the boost-invariant longitudinal impact-parameter defined as $\sigma=\frac{1}{2}b^-P^+$. The Fourier transformation of the GPDs with respect to $\xi$ provides the distributions  in the boost-invariant longitudinal space $\sigma$. Notably, the DVCS amplitude in $\sigma$ space gives an interesting diffraction pattern~\cite{Brodsky:2006in,Brodsky:2006ku}, analogous to the diffractive scattering of a wave in optics. Hence, exploring the GPDs in the longitudinal impact parameter space is intriguing. The GPDs in the longitudinal-impact parameter space are defined as:
\begin{equation}
\begin{aligned}
f(x,\sigma,t)&=\int_0^{\xi_f}\frac{d\xi}{2\pi}e^{i\xi P^+b^-/2}F(x,\xi,t)\\
&=\int_0^{\xi_f}\frac{d\xi}{2\pi}e^{i\xi\sigma}F(x,\xi,t),
\end{aligned}
\label{LFT}
\end{equation}
where $F$ stands for any one of our GPDs. The upper integration limit, denoted as $\xi_{f}$, serves as the equivalent of the slit width, representing a crucial condition for the occurrence of the diffraction pattern.
Since, we consider the region $\xi<x<1$, the upper limit of the integration is given by $\xi_f=\rm{min}\{x,\xi_{max}\}$, where for a fixed value of $- t$, the maximum value of $\xi$ is given by~\cite{Brodsky:2000xy}
\be
\xi_{\rm max}=\frac{1}{\left[1+\frac{4 M^2}{(-t)}\right]^{\frac{1}{2}}}.
\ee

\begin{figure*}[htp]
\includegraphics[scale=.35]{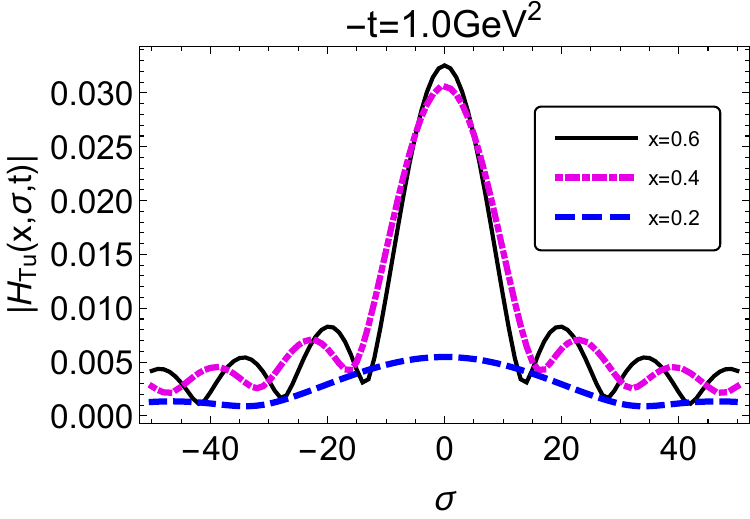}
\includegraphics[scale=.35]{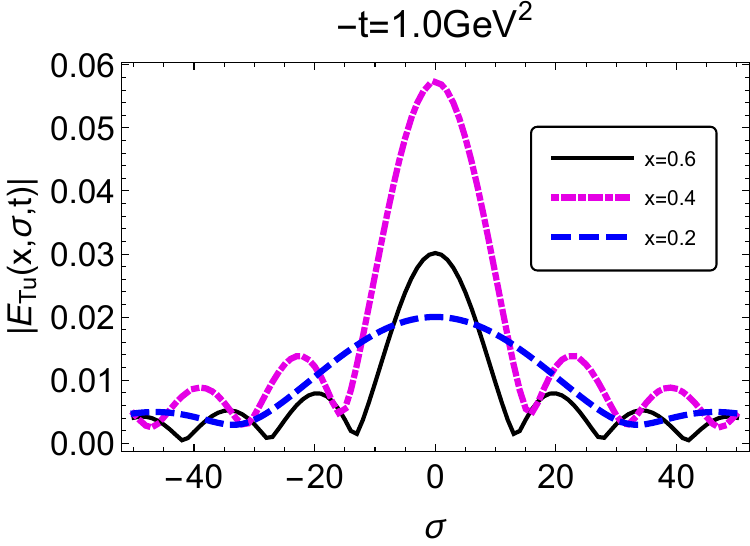} 
\includegraphics[scale=.35]{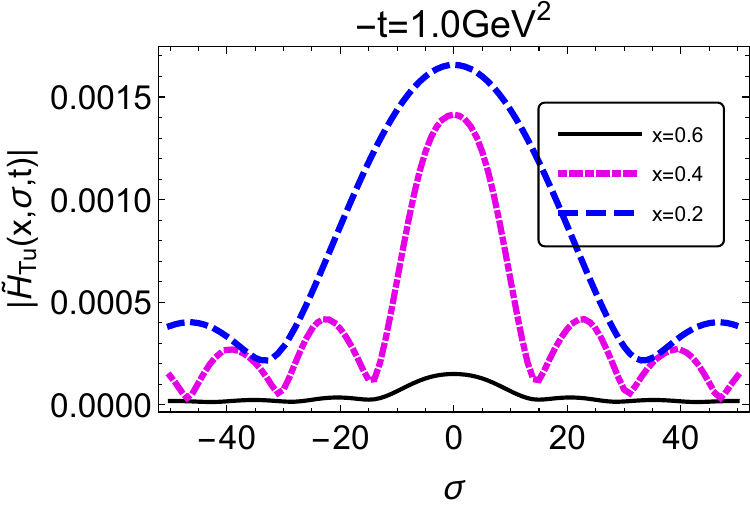} 
\includegraphics[scale=.35]{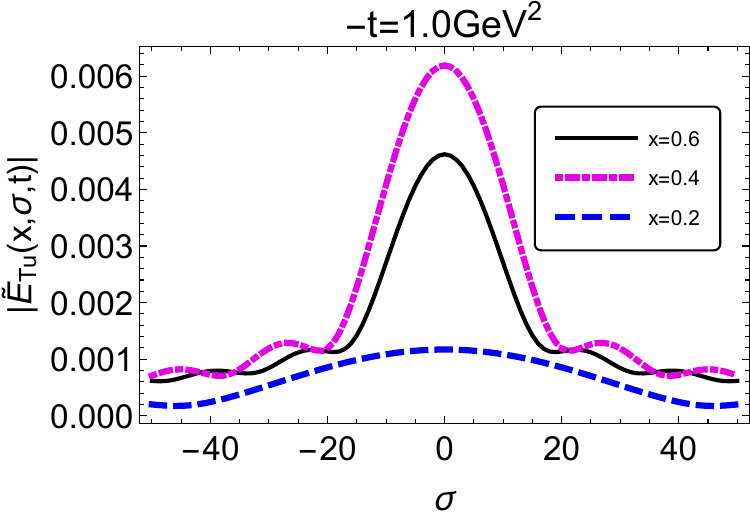} \\
\includegraphics[scale=.35]{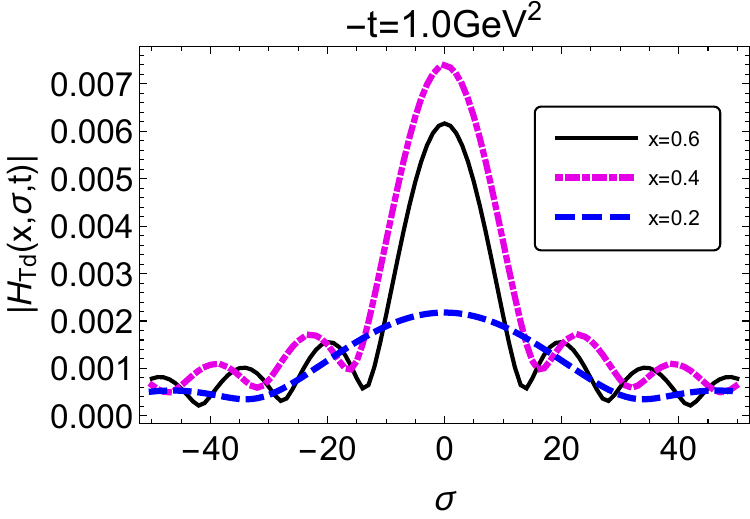}
\includegraphics[scale=.35]{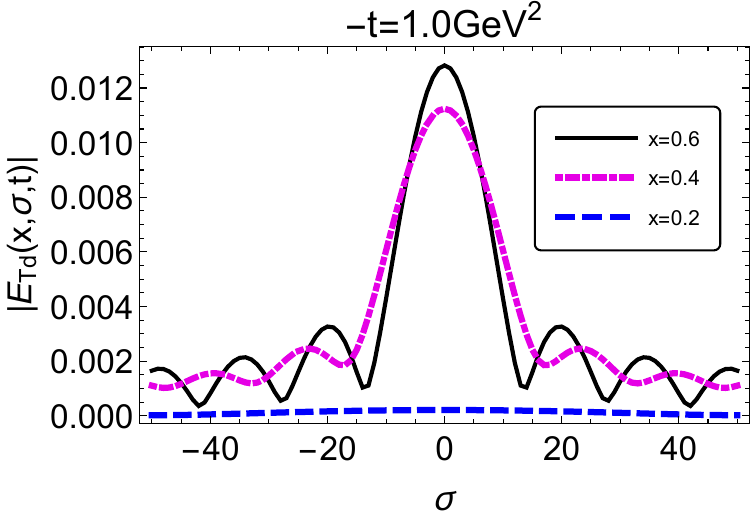} 
\includegraphics[scale=.35]{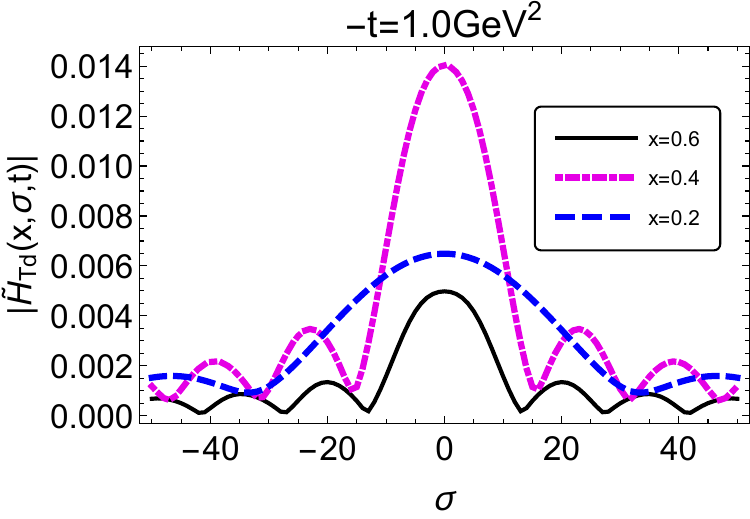} 
\includegraphics[scale=.35]{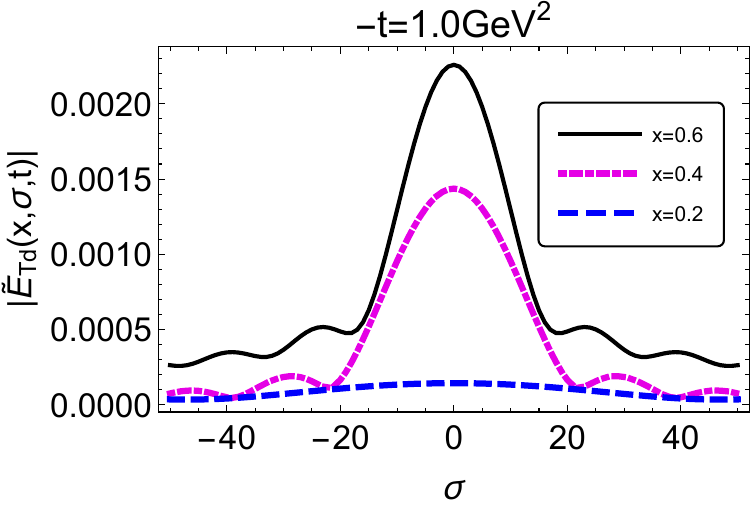}
\caption{\label{chiral_odd_xs} The chiral-odd GPDs in boost-invariant longitudinal position space as functions of $\sigma$ for  $x=\{0.2,\,0.4,\,0.6\}$ and fixed $t=-1.0$ GeV$^2$. The upper (lower) panel is for the $u \,(d)$  quark.  Left to right panels represent the GPDs $H_T,\, E_T,\, \tilde{H}_T$, and  $\tilde{E}_T$, respectively.}
\end{figure*}

Our results for the modulus of the chiral-even and chiral-odd GPDs in longitudinal position space are shown in Figs.~\ref{chiral_even_xs} and \ref{chiral_odd_xs}, respectively for three different values of $x=\{0.2,\,0.4,\,0.6\}$ and fixed value of $t=-1.0$ GeV$^2$. The  value of $-t$ corresponds to the value of $\xi_{\rm max} \approx 0.47$, which reflects the upper limit of the $\xi$ integration in Eq.~(\ref{LFT}), $\xi_f=0.2$ and $0.4$ for $x=0.2$ and $0.4$, respectively, while for $x=0.6$ the integration limit is $\xi_f=0.47$ since for this case $x>\xi_{\rm max}$. 

Our results exhibit oscillatory behavior akin to the diffraction pattern observed in a single-slit experiment in optics. The width of the principal maxima in this diffraction pattern is inversely proportional to the slit width. The finite size of $\xi$ in the Fourier transformation described by Eq.~\eqref{LFT} is responsible for generating the diffraction pattern, with $\xi_f$ serving as the equivalent of the slit width in the single-slit experiment. It is important to note that the Fourier transform of any arbitrary function with a finite range of $\xi$ does not always yield a diffraction pattern~\cite{Manohar:2010zm}. For example, $\tilde{E}_T(x,\sigma,t)$ does not exhibit the same pattern. This is due to the distinctly different nature of $\tilde{E}_T(x,\xi,t)$ with $\xi$ compared to the other
GPDs. We observe that as $\xi_f$ increases, the width of the principle maxima consequently decreases. In other words, the position of the first minima shifts towards the center with increasing $\xi_f$. We note that a similar diffraction pattern in $\sigma$-space has also been noticed in the DVCS amplitude~\cite{Brodsky:2006in,Brodsky:2006ku}, the coordinate-space parton density~\cite{Miller:2019ysh}, the GPDs extracted in different
phenomenological models ~\cite{Chakrabarti:2008mw,Manohar:2010zm,Kumar:2015fta,Mondal:2015uha,Chakrabarti:2015ama,Mondal:2017wbf,Kaur:2018ewq} as well as in more general distributions, the Wigner distributions~\cite{Maji:2022tog,Ojha:2022fls}. Notably, all the distributions in the boost-invariant longitudinal space exhibit a long-distance tail, as highlighted in Refs.~\cite{Miller:2019ysh,Weller:2021wog}.
\section{Conclusions}
The study of skewed GPDs is of interest because it can be connected to exclusive scattering cross sections
and can be measured through the experiments at the EIC and EicC.
BLFQ has emerged as a nonperturbative tool for solving many-body bound state problem in quantum field theory. In this work, we  calculated all the leading-twist skewness-dependent quark GPDs for the proton from its LFWFs within the framework of BLFQ. These wave functions were obtained from the eigenvectors of an effective light-front Hamiltonian within the leading Fock sector, which includes a three-dimensional confining potential and a one-gluon exchange interaction with fixed coupling and  numerical cutoffs. We presented
the GPDs considering the DGLAP region, i.e., for $x>\xi$ and found that the qualitative behavior of the GPDs in our BLFQ approach exhibits similarities to other phenomenological models.

The GPDs in longitudinal impact-parameter space provide a unique way to visualize the structure
of the proton.
We obtained the GPDs in the boost invariant longitudinal position space, $\sigma=\frac{1}{2}b^-P^+$, by performing the Fourier transform of the skewed GPDs with respect to $\xi$. We observed that the GPDs in longitudinal position space for a fixed $x$ and $-t$ exhibit a diffraction pattern.
In optics, a comparable diffraction pattern is observed in a single-slit experiment, where the width of the central maxima is inversely proportional to the slit width. Our findings are analogous to the diffractive scattering of waves in optics, with the finiteness of the $\xi$ integration (represented by the upper limit, $\xi_f$) playing the role of the slit width. However, the diffraction pattern is not solely attributed to the finiteness of $\xi$ integration; the functional behaviors of the GPDs also play important roles for this phenomenon. A similar diffraction pattern has also been identified in various observables, such as the DVCS amplitude, the parton density, and Wigner distributions in longitudinal position space.

Our ongoing efforts to extend to higher Fock sectors, especially the  $|qqqq\bar{q}\rangle$ and $|qqqg\rangle$ Fock sectors, will allow us to evaluate the skewed GPDs in the ERBL region, i.e., for $x<\xi$ as well as to investigate the gluon and sea-quark GPDs. Our approach can also be utilized to calculate higher-twist GPDs~\cite{Zhang:2023xfe}.

\section*{Acknowledgements}
CM thanks the Chinese Academy of Sciences President's International Fellowship Initiative for the support via Grants No.~2021PM0023. CM is also supported by new faculty start up funding by the Institute of Modern Physics, Chinese Academy of Sciences, Grant No.~E129952YR0. XZ is supported by new faculty startup funding by the Institute of Modern Physics, Chinese Academy of Sciences, by Key
Research Program of Frontier Sciences, Chinese Academy of Sciences, Grant No. ZDB-SLY-7020, by the Natural Science Foundation of Gansu Province, China, Grant No. 20JR10RA067, by the Foundation for Key Talents of Gansu Province, by the Central Funds Guiding the Local
Science and Technology Development of Gansu Province, Grant No. 22ZY1QA006, by international partnership
program of the Chinese Academy of Sciences, Grant No. 016GJHZ2022103FN, by the National Natural Science Foundation of China under Grant No. 12375143, by National Key R\&D Program of China, Grant
No. 2023YFA1606903, and by the Strategic Priority
Research Program of the Chinese Academy of Sciences,
Grant No. XDB34000000. JPV is supported by the U.S.
Department of Energy under Grant No. DE-SC0023692. A portion of the computational resources were also provided by Gansu Computing Center. This research is supported by Gansu International Collaboration and Talents Recruitment Base of Particle
Physics (2023–2027) and the International Partnership Program of Chinese Academy of Sciences, Grant No. 016GJHZ2022103FN.

\biboptions{sort&compress}
\bibliographystyle{elsarticle-num}
 \bibliographystyle{apsrev4-1}
\bibliography{ref.bib}

\end{document}